\documentclass[preprint,showpacs,showkeys,preprintnumbers,aps]{revtex4}

\usepackage{graphicx}

\begin{document}

\preprint{Fecher et al., HEHRPES of Co$_2$Mn$_{1-x}$Fe$_x$Si.}

\title{High energy, high resolution photoelectron spectroscopy of Co$_2$Mn$_{1-x}$Fe$_x$Si.}

\author{Gerhard H. Fecher, Benjamin Balke, Siham Ouardi, and Claudia Felser}
\email{fecher@uni-mainz.de}
\affiliation{Institut f\"ur Anorganische und Analytische Chemie, \\
Johannes Gutenberg - Universit\"at, D-55099 Mainz, Germany.}

\author{Eiji Ikenaga, Jung-Jin Kim, Shigenori Ueda, and Keisuke Kobayashi}
\address{Japan Synchrotron Radiation Research Institute (SPring-8/JASRI), \\
         Kouto 1-1-1, Sayo-cho, Sayou-gun,
         Hyogo, 679-5198, Japan}
         
\date{\today}

\begin{abstract}

This work reports on high resolution photoelectron spectroscopy for
the valence band of Co$_2$Mn$_{1-x}$Fe$_x$Si ($x=0,0.5,1$) excited by
photons of about 8~keV energy. The measurements show a good agreement
to calculations of the electronic structure using the LDA$+U$ scheme.
It is shown that the high energy spectra reveal the bulk electronic
structure better compared to low energy XPS spectra. The high
resolution measurements of the valence band close to the Fermi energy
indicate the existence of the gap in the minority states for all three
alloys.
\end{abstract}



\maketitle

\section{Introduction}

Electronic devices exploiting the spin of an electron (spintronics
\cite{WAB01}) have attracted great scientific interest in particular
for magneto-electronics \cite{Pri98}. The basic element is a
ferromagnetic electrode providing a spin polarised electrical current.
Materials with a complete spin polarisation would be most desirable,
i.e. a metal for spin up and an insulator for spin down electrons. Such
materials are called half-metallic ferromagnets \cite{GME83,CVB02}.
K{\"u}bler {\it et al} \cite{KWS83} recognised that the minority-spin
state density at the Fermi energy  nearly vanishes in the Heusler
compounds Co$_2$MnAl and Co$_2$MnSn. The authors concluded that this
should lead to peculiar transport properties in these compounds because
only the majority density contributes. The Heusler alloy Co$_2$MnSi has
attracted particular interest because it is predicted to have a large
minority spin band gap of 0.4~eV and, at 985~K, has one of the highest
Curie temperatures, among the known Heusler compounds
\cite{FSI90,BNW00}. An even higher Curie temperature (1100~K)
accompanied by a large magnetic moment (6~$\mu_B$) was found in recent
investigations of Co$_2$FeSi \cite{WFK05,WFK06a,KFF06}. The end members
of the series Co$_2$Mn$_{1-x}$Fe$_x$Si, that are the purely Mn or Fe
containing compounds, have been used for fabrication of magnetic tunnel
junctions \cite{IOM06,OSN06}. The tunnel magneto-resistance ratios of
159\% in the Mn compound at low temperature and 41\% in the Fe compound
at room temperature suggest that still an improvement in the materials
is necessary for successful use in devices.

It was recently shown that the complete substitutional series
Co$_2$Mn$_{1-x}$Fe$_x$Si ($0 \leq x \leq 1$) orders in the Heusler type
$L2_1$ structure \cite{BFK06}. Magneto-structural investigations using
$^{57}$Fe M{\"o\ss}bauer spectroscopy confirmed the high degree of
structural order. A structural phase transitions at about 1025~K was
detected by means of differential scanning calorimetry. Low temperature
magnetometry confirmed a Slater-Pauling like behaviour \cite{FKW06} of
the complete series of alloys with the magnetic moment increasing
linearly from 5~$\mu_B$ to 6~$\mu_B$ with increasing iron content $x$
from~0 to~1.

Photoelectron spectroscopy is one of the best suited techniques to study
the occupied electronic structure of materials. Low kinetic energies
result in a small electron mean free path of less than 5.2~{\AA} at
kinetic energies below 100~eV. That is, the escape depth of the
electrons is less than one cubic cell of the investigated compounds and
mainly the surface will contribute to the intensity if using UV light
for excitation. This very high surface sensitivity may be able to
explain the quite low spin polarisation of photoelectrons emerging from
single crystalline Co$_2$MnSi films at the Fermi level of only 12~{\%}
\cite{WPK05b}. Wang {\it et al} \cite{WPK05a,WPK05b} assumed that
partial chemical disorder was responsible for this discrepancy with the
theoretical predictions.

In the common X-ray photoelectron spectroscopy (XPS) one uses medium
energies for excitation being provided by Al-K$_\alpha$ or
Mg-K$_\alpha$ sources resulting in a probing depth of about 24~{\AA} at
1.2~keV. The situation becomes much better at higher energies. In hard
X-ray photoelectron spectroscopy with excitation energies of about
8~keV one will reach a high bulk sensitivity with an escape depth being
larger than 115~{\AA} (corresponding to 20 cubic cells). Lindau {\it et
al} \cite{LPD74} demonstrated in 1974 the possibility of high energy
photoemission with energies up to 8~keV, however, no further attention
was devoted to such experiments for many years. High energy
photoemission (at about 15~keV excitation energy) was also performed as
early as 1989 \cite{Mei89} using a $^{57}$Co M{\"o\ss}bauer
$\gamma$-source for excitation, however, with very low resolution only.
Nowadays, high energy excitation and analysis of the electrons become
easily feasible due to the development of high intense sources
(insertion devices at synchrotron facilities) and multi-channel
electron detection. Thus, high energy photoemission spectroscopy was
recently introduced by several groups
\cite{KYT03,SS04,TKC04,Kob05,PCC05,TSC05} as a bulk sensitive probe of
the electronic structure in complex materials.

In Ref.~\cite{WFK06b} it was demonstrated that high energy
photoelectron spectroscopy is a useful tool to study the electronic
structure of complex Heusler alloys using the example of
Co$_2$Cr$_{0.6}$Fe$_{0.4}$Al. In that work photon energies of 3.5~keV
were used for excitation, and compared to resonant, medium energy
(0.5-0.8~keV) excitation at the $L_{3,2}$ edges of the contributing
$3d$-transition metals. In the present work, an excitation energy of
$h\nu\ = 8$~keV was used to study the density of states of
Co$_2$Mn$_{1-x}$Fe$_x$Si with $x=0,0.5,1$.

\section{Experimental and computational details}
\label{sec:ED}

Co$_2$Mn$_{1-x}$Fe$_x$Si samples were prepared by arc melting of
stoichiometric amounts of the constituents in an argon atmosphere at
10$^{-4}$~mbar. Care was taken to avoid oxygen contamination. This was
ensured by evaporating Ti inside of the vacuum chamber before melting
the compound as well as by additional purifying of the process gas.
After cooling of the resulting polycrystalline ingots, they were
annealed in an evacuated quartz tube for 21~days. Rods with a dimension
of $(1\times 1\times 5)$~mm$^3$ were cut from the ingots by
spark-erosion for experiments on in-situ fractured samples. Flat discs
with about 10~mm diameter and 1~mm thickness were cut and polished for
spectroscopic investigations. Further experimental details and results
of the structural and magnetic properties are reported in
Ref.~\cite{BFK06,KEB06}.

X-ray photoemission spectroscopy was used to verify the composition
and to check the cleanliness of the samples. After removal of the
native oxide from the polished surfaces by Ar$^+$ ion bombardment, no
impurities were detected with XPS. The ESCALAB Mk~II (VG) was also used
to take valence band spectra at 1253.6~eV (Mg~K$_\alpha$ with a natural
line width of 0.69~eV \cite{BSh76}) for comparison. For this purpose,
the slits and pass energy were set for an analyser-resolution of
200~meV. The energy was calibrated at the Au 4f$_{7/2}$ emission line.

The electronic structure was explored experimentally by means of high
energy X-ray photoemission spectroscopy (HXPS). The measurements were
performed at the beamline BL47XU of SPring~8 (Hyogo,
Japan). The photons are produced by means of a 140-pole in-vacuum
undulator and are further monochromatised by two double-crystal
monochromators. The first monochromator uses Si(111) crystals and the
second a Si(111) channel-cut crystal with 444 reflections (for 8~keV
X-rays). The energy of the photoemitted electrons is analysed using a
Gammadata - Scienta R~4000-10kV electron spectrometer. The ultimate
resolution of the set up (monochromator plus analyser at 50~eV pass
energy using a 200~$\mu$m slit) is 83.5~meV at 7935.099~eV photon
energy as employed for the reported experiments. Under the present
experimental conditions an overall resolution of 250~meV has been
reached for the valence band spectra and 130~meV for the spectra taken
at the Fermi energy. All values concerning the resolution are
determined from the Fermi-edge of an Au sample. Due to the low
cross-section of the valence states from the investigated compounds,
the spectra had to be taken with $E_{pass}=200$~eV and a 500~$\mu$m
(200~$\mu$m) slit for a good signal to noise ratio. The polycrystalline
samples have been fractured in-situ before taking the spectra to remove
the native oxide layer. Core-level spectra have been taken to check the
cleanliness of the samples. No traces of impurities were found. The
valence band spectra were collected over up to 4~h at about 100~mA
electron current in the storage ring in the top-up mode. All
measurements have been taken at a sample temperature of 20~K.

The self-consistent electronic structure calculations have been carried
out using the full potential linearised augmented plane wave method
(FLAPW) as provided by Wien2k \cite{BSM01}. The exchange-correlation
functional was taken within the generalised gradient approximation
(GGA) as introduced by Perdew {\it et al} \cite{PBE96}. A
$25\times25\times25$ mesh has been used for integration, resulting in
455 $k$-points in the irreducible wedge of the Brillouin zone of the
primitive cell of the cubic compounds. All muffin tin radii have been
set as nearly touching spheres with $r_{MT}=2.29 a_{0B}$ for the $3d$
elements and $2.15 a_{0B}$ for Si ($a_{0B} = 0.529177$~{\AA}). A
structural optimisation for the compounds showed that the calculated
lattice parameters deviate from the experimental ones only marginally
\cite{KFF06}. Further details of the calculations and their results are
reported in Reference~\cite{KFF06,BFK06}.

Partial cross sections have been calculated for better comparison of
the calculated electronic structure with the valence band photoemission
spectra. The orbital momentum and site resolved cross sections were
calculated for atomic valence states using a modified relativistic
Dirac-solver based on the computer programs of Salvat and Mayol
\cite{SMa91,SMa93}. The radial integrals for the various transitions
($s \rightarrow p$, $p \rightarrow s,d$, and $d \rightarrow p,f$) have
been computed using the dipole length-form. In addition, the electron
mean free path was calculated using the Tanuma-Powell-Penn (TPP-2M)
equations \cite{TPP93}.

\section{Results and Discussion}

\begin{figure}
\includegraphics[width=6.5cm]{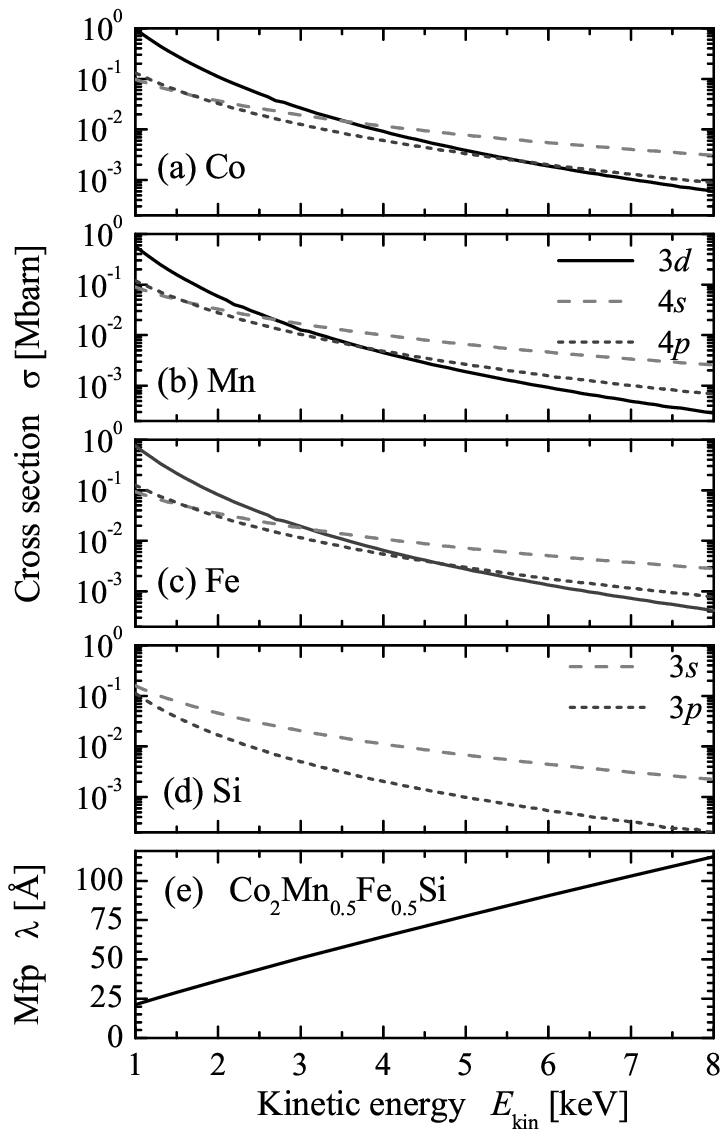}
\caption{Cross sections and electron mean free path. \newline
         The atomic, partial cross sections of the contributing elements
         are displayed in (a)-(d) and the electron mean free path (Mfp)
         as calculated for Co$_2$Mn$_{0.5}$Fe$_{0.5}$Si in (e).}
\label{fig_1}
\end{figure}

Figure \ref{fig_1} displays the calculated energy dependence of the
atomic and orbital resolved cross sections together with the electron
mean free path in Co$_2$Mn$_{0.5}$Fe$_{0.5}$Si. The shown data cover
the range of kinetic energies observed in XPS and HXPS. In this energy
range, the electron mean free path (Fig.~\ref{fig_1}(e)) varies nearly
linearly with increasing kinetic energy and covers a range of escape
depths from about 5 to 20 cubic Heusler cells. The calculated electron
mean free path in the pure Mn or Fe containing compounds is very close
to the one of the mixed compound as the physical properties (density,
number and kind of valence electrons) entering the TPP-2M equations are
almost the same.

At low energies (1~keV), like used in regular XPS, the partial cross
section of the valence states is dominated by the $d$-states of the
transition metal elements. This cross section decreases faster with
increasing energy compared to that of the $s$ or $p$ states. At high
energies (8~keV) the strongest cross sections is observed for the
$s$-states, because the cross section of $p$ states also drops down
faster with energy compared to $s$-states. Thus one expects a
pronounced change of the photoemission intensities if going from XPS
(1.2~keV) to HXPS (8~keV).

\begin{figure}
\includegraphics[width=9cm]{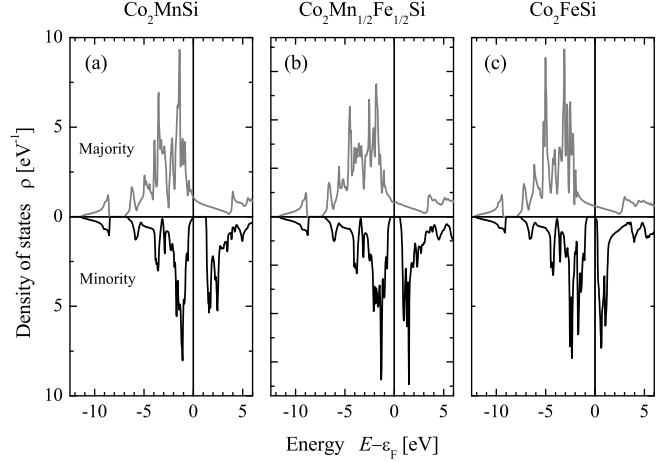}
\caption{Spin resolved density of states of
         Co$_2$Mn$_{1-x}$Fe$_{x}$Si for $x=0,1/2,$~and~$1$.}
\label{fig_2}
\end{figure}

Figure \ref{fig_2} displays the calculated, spin resolved density of
states for the investigated compounds. All three compounds,
Co$_2$Mn$_{1-x}$Fe$_{x}$Si with $x=0,1/2,$ and $1$, exhibit a clear gap
in the minority density of states, that is they are half-metallic
ferromagnets. The gap has the result that the calculated spin magnetic
moments of Co$_2$MnSi and CoFeSi have integer values of 5~$\mu_B$ and
6~$\mu_B$, respectively. The magnetic moment of
Co$_2$Mn$_{1/2}$Fe$_{1/2}$Si is 5.5~$\mu_B$. All those values are in
perfect agreement with the experimental values \cite{BFK06} and the
Slater-Pauling rule \cite{FKW06}. The majority spin density reveals
that the $d$-states emerging from flat majority bands are shifted away
from the Fermi energy $\epsilon_F$ with increasing Fe content. At the
same time the Fermi energy moves from near the top of the minority
valence states towards the bottom of the minority conduction bands. It
is clearly visible that the gap in the minority states is defined by
regions of high density emerging from flat $d$-bands. At the same time,
the majority spin density contributes only few states in the region
close to the Fermi energy. This behaviour is typical for Heusler
compounds with high magnetic moments. More details of the electronic
structure are reported in Refs.~\cite{KFF06,BFK06}.

\begin{figure}
\includegraphics[width=9cm]{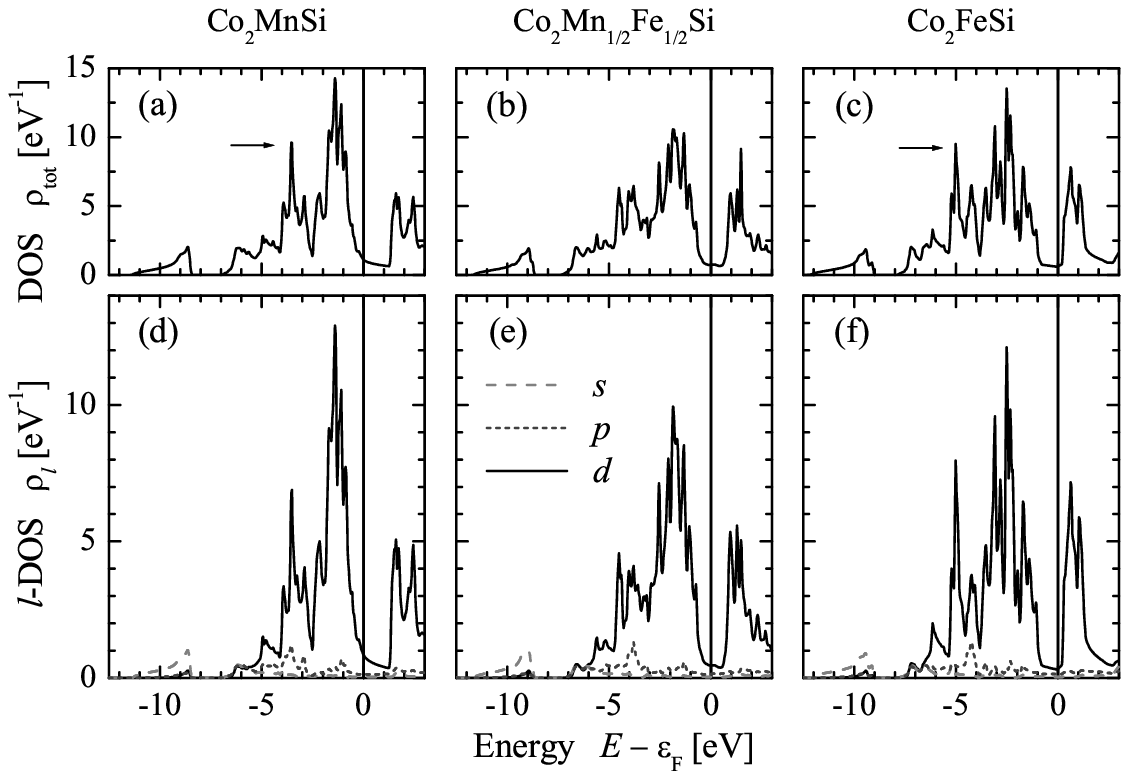}
\caption{Spin averaged density of states of Co$_2$Mn$_{1-x}$Fe$_{x}$Si. \newline
         The total, spin averaged density of states $\rho_{tot}$ is shown in (a)~-~(c) for $x=0,1/2$, and~$1$.
         The corresponding orbital momentum resolved $l$-DOS $\rho_{l}$ is displayed in {(d)~-~(f)}.}
\label{fig_3}
\end{figure}

The particular shape of the spin densities close to $\epsilon_F$ - a
low density emerging from majority states surrounded by regions of high
minority density - should be easily detected in valence band
photoemission and thus may be a good indicator for the existence of the
gap, even without spin analysis. Figures \ref{fig_3}(a)-(c) show the
total, spin integrated density of states. The low density of states in
the vicinity of the Fermi energy is clearly visible.

Additionally, the orbital momentum resolved density of states ($l$-DOS)
is shown in Figure~\ref{fig_3}(d)-(f). The partial DOS of the
interstitial can not be extracted from the calculations in an
$l$-resolved way and therefore is not included. The $l$-DOS for higher
angular momenta ($l$) is omitted as they contribute only very few to
the total density of states. From Figure~\ref{fig_3}(d)-(f) it follows
that the $s$-states are mainly found at below 8.5~eV below
$\epsilon_F$. The density at the Fermi energy is dominated by
$d$-states. Although the density of $d$-states is already low at
$\epsilon_F$, the $s$-like density is still at least one order of
magnitude smaller. The density related to $p$-states is found from the
bottom of the $d$-bands up to about 1~eV below $\epsilon_F$. This
behaviour reflects the interaction between transition metal $d$-states
with Si $p$-states. For better comparison with the measured
photoemission spectra, the $l$-DOS has been weighted in the following
by the partial cross sections. The sum of the weighted $l$-DOS has been
additionally convoluted by the Fermi-Dirac distribution (300~K for XPS;
20~K for HXPS) and afterwards has been broadened by Gaussians of 0.7~eV
(XPS) or 0.27~eV (HXPS) width to account roughly for the experimental
resolution at excitation energies of 1.2~keV or 8~keV, respectively.

\begin{figure}
\includegraphics[width=9cm]{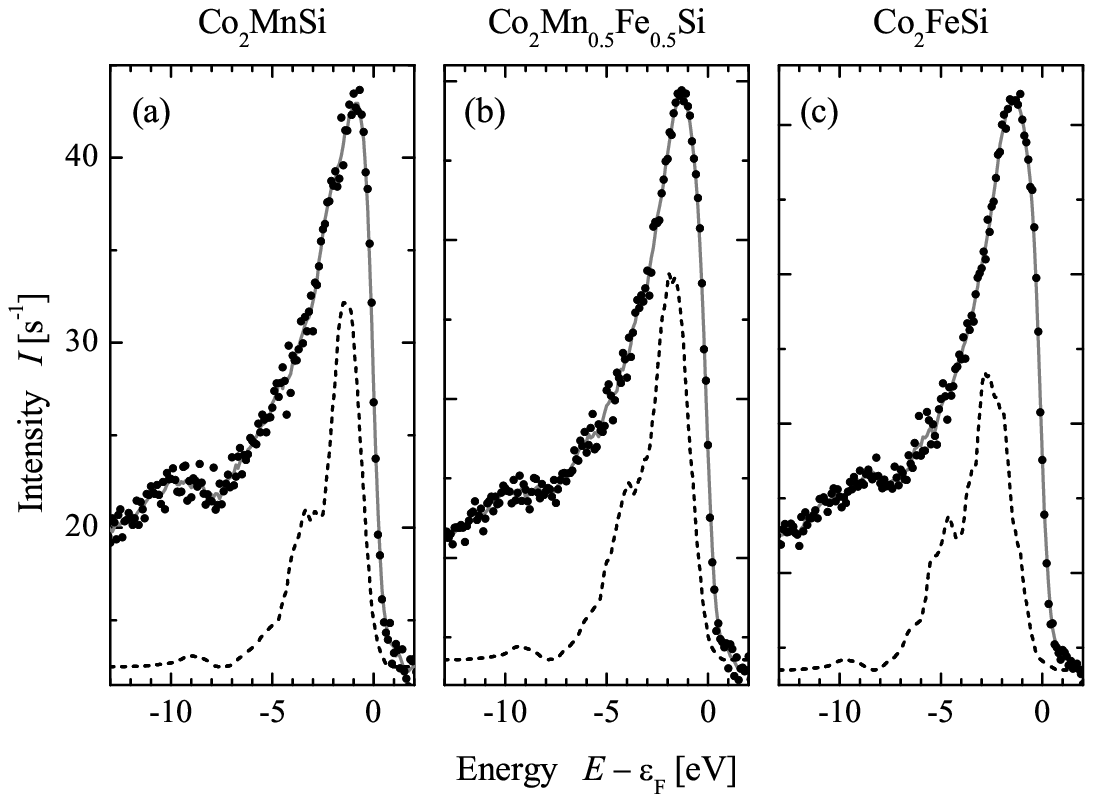}
\caption{Low energy valence band spectra of Co$_2$Mn$_{1-x}$Fe$_x$Si. \newline
         (a)~-~(c) display the XPS valence band spectra for $x=0,0.5$, and~$1$.
         The DOS - convoluted by the Fermi-Dirac distribution and
         weighted by the partial cross sections - is shown as dashed line.
         The spectra were excited by Mg~K$_{\alpha}$ radiation (1.254~keV).}
\label{fig_4}
\end{figure}

The valence band spectra excited by Mg~K$_{\alpha}$ radiation are shown
in Figure~\ref{fig_4}. All three spectra show a high intensity close to
the Fermi energy. The low lying $sp$ band at 8~eV to 11~eV below
$\epsilon_F$ is only weakly revealed. There is also no particular
structure seen close to $\epsilon_F$. However, a closer inspection of
the maximum intensity of the emission exhibits that it is shifted away
from the Fermi energy in the Fe containing alloys (b,c). The maximum
intensity is found at -1.3~eV, -1.46~eV, and -1.55~eV for $x=0$, 0.5,
and 1, respectively. This observation is in rough agreement with the
density of states where the maximum of the majority density (as well as
the total DOS) is also shifted to higher binding energies with
increasing Fe content. It is also seen in the DOS weighted by the
partial cross sections (dashed lines in Fig.~\ref{fig_4}) where the
maximum of the intensity shifts from -1.48~eV for $x=0$ to -2.85~eV for
$x=1$. Both values are in the calculations larger than those observed
in the spectra. In particular, this suggests that the calculations
shift the $d$-states in Co$_2$FeSi to much below the Fermi energy, what
may be prevented by using different values of $U_{eff}$ such that the
Fermi energy comes closer to the top of the minority valence band
\cite{KFF06}.

\begin{figure}
\includegraphics[width=9cm]{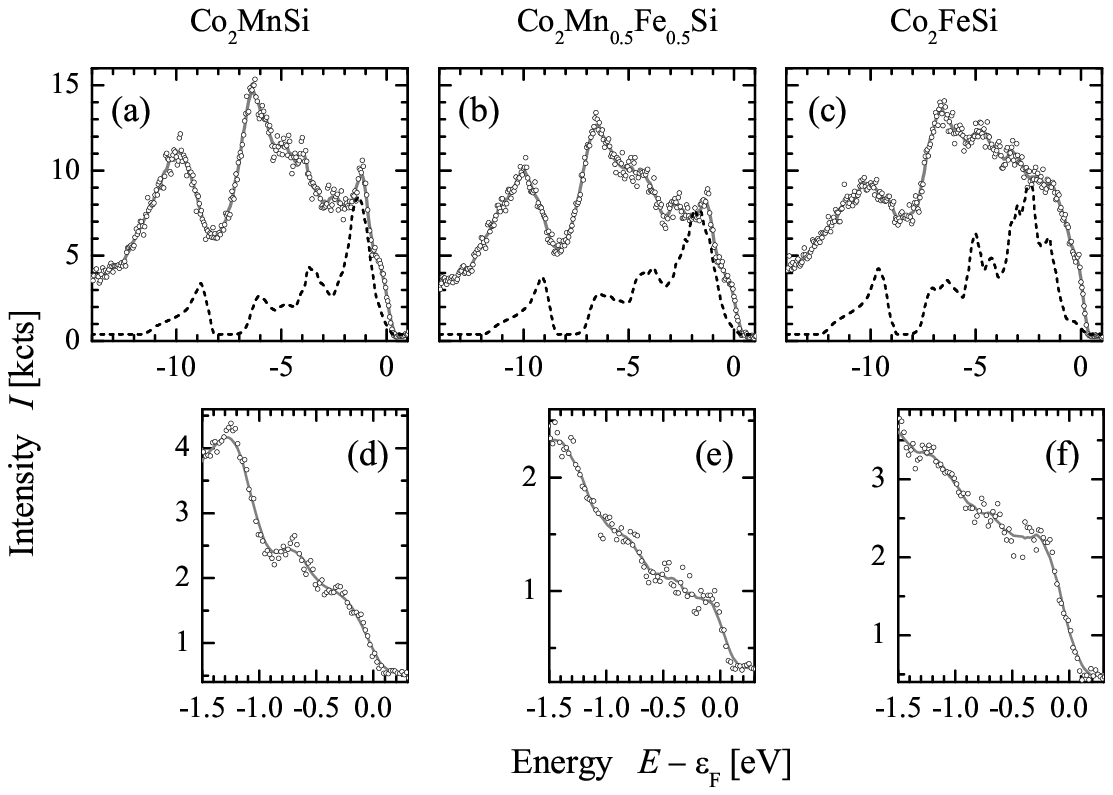}
\caption{High energy valence band spectra of Co$_2$Mn$_{1-x}$Fe$_x$Si. \newline
         The complete valence band is shown in panels (a)~-~(c) for $x=0,0.5$, and~$1$.
         The DOS - convoluted by the Fermi-Dirac distribution and
         weighted by the partial cross sections - is shown as dashed line.
         High resolution valence band spectra taken close to the
         Fermi energy are displayed in (d)~-~(f).
         The spectra were excited by synchrotron radiation with $h\nu=7.939$~keV.}
\label{fig_5}
\end{figure}

The results from high energy photoemission are shown in
Figure~\ref{fig_5}. The spectra of all three compounds reveal clearly
the low lying $s$-states at about -11~eV to -9~eV below the Fermi
energy, in well agreement to the calculated DOS. These low lying bands
are separated from the high lying $d$-states by the Heusler-typical
hybridisation gap being clearly resolved in the spectra as well as the
calculated DOS. The size of this gap amounts typically to $\Delta
E\approx 2$~eV in Si containing compounds.

Obviously, the emission from the low lying $s$-states is pronouncedly
enhanced compared to the emission from the $d$-states. This can be
explained by a different behaviour of the cross sections of the $s$,
$p$, and $d$ states with increasing kinetic energy as was recently
demonstrated by Panaccione {\it et al} for the case of the silver
valence band \cite{PCC05}. In particular, the cross section for
$d$-states decreases faster with increasing photon energy than the one
of the $s$-states. This behaviour influences also the onset of the
$d$-bands at about -7~eV. Just at the bottom of those $d$-bands, they
are hybridised with $s, p$-like states, leading to a relatively high
intensity in this energy region.

The structure of the spectra in the range of the $d$ states agrees with
the structures observed in the total DOS. However, one has to account
not only for the experimental resolution but also for lifetime
broadening if comparing that energy range. The lowest flat band of the
majority band structure, accompanying the localised moment, results in
a sharp peak in the DOS at about -3.5~eV and -5~eV for Mn and Fe,
respectively (marked by arrows in Fig.\ref{fig_3}~(a) and~(c). These
peaks are shifted away from $\epsilon_F$ by the electron-electron
correlation in the LDA$+U$ calculation and would appear without $U$
closer to the Fermi energy. Their energetic position corresponds to
structures revealed in the measured spectra, thus they are a good proof
for the use of the LDA$+U$ scheme.

As mentioned, one also has to account for lifetime broadening. At
1.3~keV excitation, the emission is dominated by the high dense
$d$-states at about -1.4~eV. Increasing the excitation energy to 8~keV
(note the better resolution of set-up at that energy) has the result
that the intensity in this energy range becomes considerably lower. At
the same time, the emission from the remaining $d$-bands becomes
strongly enhanced. Obviously, the intensity at the Fermi energy is much
higher compared to the calculated DOS even after weighting by the
partial cross sections. As those structures in the DOS emerge rather
all from $d$-states, the transfer of the intensity maximum is not only
explained by pure cross section effects. The valence band spectra may
be seen as a convolution of the initial (bound) and final (unoccupied)
state DOS. The final state DOS is rather constant at high kinetic
energies and final state effects may play a minor rule only. Two other
weighting factors enter the DOS-convolution. The first is, as explained
above, the transition matrix element that contains both the selection
rules and the cross sections (radial matrix elements). The radial
matrix elements are partially responsible for the rearrangement of the
orbital resolved intensities as discussed above. The second weighting
factor is the complex self-energy of the photoelectron. Among other
things, it depends on the hole lifetime due to the coupling of the
photoemitted electron with the hole left behind. At low kinetic
energies, the spectra are obviously governed by the long life time of
the holes at binding energies close to $\epsilon_F$. At high kinetic
energies, where the sudden approximation is reached, the photoelectron
is not as strongly coupled to its hole and the lifetime at $\epsilon_F$
plays less a role. Not only from the mean free path but also from the
presented point of view, the high energy photoemission may help to
understand the bulk electronic structure better than using only low
energy XPS.

Most interesting is the behaviour of the calculated DOS and the
measured spectra close to $\epsilon_F$ as this might give an advice
about the gap in the minority states. The majority band structure
contributes only few states to the density at $\epsilon_F$ emerging
from strongly dispersing bands. This region of low density is enclosed
by a high density of states arising from flat bands at the upper and
lower limits of the minority band gap. The onset of the minority
valence band is clearly seen in the total DOS as well as the low
majority density at the Fermi energy. The same behaviour is principally
observed in the measured valence band spectra as displayed in
Figure~\ref{fig_5}(d)-(f). From the spectra, it can be estimated that
the Fermi energy is in all three cases at least $\approx 0.5$~eV above
the minority states with high density (see Fig.~\ref{fig_2}). This
gives strong evidence that all compounds of the
Co$_2$Mn$_{1-x}$Fe$_x$Si series exhibit really half-metallic
ferromagnetism. Actually it stays unexplained why there is a strong
enhancement of the intensity emerging from the steep majority $d$-bands
crossing the Fermi energy compared to the lower lying flat $d$-bands.
The answer will need to use more sophisticated photoemission
calculations that include the angular matrix elements to distinguish
between $e_g$ and $t_{2g}$ states. However, such programs are presently
not available for the high kinetic and photon energies as used in this
work.

The values for $U_{eff}$ used here are the borderline cases for the
half-metallic ferromagnetism over the complete series
Co$_2$Mn$_{1-x}$Fe$_x$Si. They were used as being independent of the Fe
concentration, what was suggested for Co from constrained LDA
calculations. However, the valence band spectra indicate that the Fermi
energy of both end members may fall inside of the minority gap rather
than being located at the edges of the minority gap. This situation may
be simulated by a variation of $U$. A comparison to the $U$-dependence
of the minority gap shown in Ref.:~\cite{KFF06} suggests smaller
effective Coulomb exchange parameters for the Mn rich part and larger
ones for the Fe rich part of the series. This might also explain the
non-linearity reported in Ref.~\cite{BFK06} for the hyperfine field. A
variation of those parameters for all contributing $3d$ constituents in
the calculations was omitted here because it would not bring more
insight into the nature of the problem, at present.

Overall, the measured photoelectron spectra agree well with the
calculated density of states and principally verify the use of the
LDA$+U$ scheme. In particular, the shape of the spectra close to
$\epsilon_F$ can be explained by the occurrence of a gap in the
minority states and thus points indirectly on the half-metallic state
of all three compounds investigated here by photoemission. For clarity
about the gap, spin resolved photoemission spectroscopy at high
energies would be highly desirable. However, this will make another
step of improvement of the instrumentation necessary, for both photon
sources as well as electron energy and spin analysers, as the spin
detection will need a factor of at least three to four orders of
magnitude more intensity for a single energy channel at the same
resolution as used here for the intensity spectra.

\section{Summary and Conclusions}

The quaternary Heusler compound Co$_2$Mn$_{1-x}$Fe$_x$Si was
investigated for $x=0, 0.5, 1$. All samples of the substitutional
series exhibit an $L2_1$ order, independent of the Fe concentration. In
agreement with the expectation from the Slater-Pauling curve for
half-metallic ferromagnets, the magnetic moment increases linearly with
$x$ from 5~$\mu_B$ to 6~$\mu_B$.

True bulk sensitive, high energy photoemission bearded out the
inclusion of electron-electron correlation in the calculation of the
electronic structure and gave an indirect advise on the gap in the
minority states. The valence band spectra indicate an increase of the
effective Coulomb exchange parameters with increasing Fe concentration.

\bigskip
\noindent{\bf Acknowledgment :}\newline

The authors are grateful for the support by G.~Sch{\"o}nhense and thank
S.~Wurmehl for preparing the Co$_2$FeSi sample and H.~C.~Kandpal for
help with the calculations. The synchrotron radiation experiments were
performed at the beamline BL47XU of SPring-8 with the approval of the
Japan Synchrotron Radiation Research Institute (JASRI) (proposal no.
2006A1476). This work was partially supported by a Grant-in-Aid for
Scientific Research (A) (No.15206006), and also partially supported by
a Nanotechnology Support Project, of The Ministry of Education,
Culture, Sports, Science and Technology of Japan. Financial support by
the Deutsche Forschungsgemeinschaft (project TP7 in research group FG
559) is gratefully acknowledged.

\bibliography{Fecher_JPhysD_CM}

\end{document}